\documentclass[a4paper,english,reprint, aps]{revtex4-1}
\usepackage[T1]{fontenc}
\usepackage[utf8]{inputenc}
\usepackage{amsmath}
\usepackage{amssymb}
\usepackage{esint}
\usepackage{graphicx}



\usepackage{babel}
\begin{document}
\title{Reconstruction of tunnel exit time and exit momentum in strong field
ionization, based on phase space methods}
\author{Szabolcs Hack}
\email{Szabolcs.Hack@eli-alps.hu}

\affiliation{ELI-ALPS, ELI-HU Non-Profit Ltd., H-6720 Szeged, Dugonics t\'{e}r 13,
Hungary}
\affiliation{Department of Theoretical Physics, University of Szeged, H-6720 Szeged,
Tisza L. krt. 84-86, Hungary }
\author{Szil\'{a}rd Majorosi}
\affiliation{Department of Theoretical Physics, University of Szeged, H-6720 Szeged,
Tisza L. krt. 84-86, Hungary}
\author{Mih\'{a}ly Benedict}
\affiliation{Department of Theoretical Physics, University of Szeged, H-6720 Szeged,
Tisza L. krt. 84-86, Hungary}
\author{Attila Czirj\'{a}k}
\email{czirjak@physx.u-szeged.hu}
\affiliation{ELI-ALPS, ELI-HU Non-Profit Ltd., H-6720 Szeged, Dugonics t\'{e}r 13,
Hungary}
\affiliation{Department of Theoretical Physics, University of Szeged, H-6720 Szeged,
Tisza L. krt. 84-86, Hungary }
\date{\today}
\maketitle

\section*{Introduction}

Strong field ionization of atoms plays a fundamental role in attosecond
physics \citep{Krausz_Ivanov}: a suitably strong laser pulse enables
an electron to leave (usually by tunneling) from its atomic bound
state into the continuum, which is the first step of the very successful
three-step model underlying our understanding of high-order harmonic
generation and attosecond metrology. Currently, the problems of tunneling
time \citep{Landsman_Keller_2015} and exit momentum in tunnel ionization
are of fundamental importance regarding both quantum theory and the
interpretation of experimental results in attosecond physics. In the
last few years, several groups published relevant experimental results
\citep{Eckle_Science_2008,shafir_nature_2012,Landsman_Keller_2014,Camus_PRL_2017},
typically with the attoclock method. However, the interpretation
of these measurements is usually difficult and there are also controversial
results. 

Isolated attosecond pulses can be generated by linearly polarized
few- or single-cycle laser pulses if the electric field of the laser
pulse exceeds a threshold value only during the half-cycle containing
the peak \cite{lewenstein1994hhgtheory}. Tunnel ionization is then possible practically only
during this half-cycle. However, tunnel ionization produces an electron
wavepacket which is blurred in space and time, thus attributing a
classical particle to this wavepacket is rather ambiguous. Some of
the electrons liberated by tunnel ionization ultimately leave the
parent ion and can be detected later, while other electrons return
to the parent ion and either rescatter on it or recombine with it.
In this latter case, the emission of a high-energy photon with a proper
phase is possible which contributes to the resulting macroscopic isolated
attosecond pulse. Due to this probabilistic nature of this quantum
process, the relative timing of the driving laser pulse and the generated
attosecond pulse (\quotedblbase time zero\textquotedblright ) a priori
can not be determined more accurately than a few tens of attoseconds,
which is leads to problems in the interpretation of experimental results
in attosecond physics. In this contribution, we analyze tunnel ionization
of a single atom based on the Wigner function over the classical phase
space which inspires improved classical electron trajectories: these
start with exit momenta based on the quantum momentum function and
correspond very well to the subsequent quantum evolution.

\section*{Theoretical model and numerical solution}

We work in the framework of a simple model: dipole approximation for
the interaction of a single active electron atom with the classical
electromagnetic field in the length gauge. We consider a near-infrared single-cycle laser pulse with sine-squared
envelope, linearly
polarized along the $z$-direction, which excites the electron from the ground state of the
bounding atomic Coulomb potential. We solve the 3D time-dependent
Schrödinger equation for the relative motion numerically, in cylindrical
coordinates: 
\begin{eqnarray}
i\hbar\frac{\partial}{\partial t}\Psi= && \left[-\frac{\hbar^{2}}{2m}\left(\frac{\partial^{2}}{\partial^{2}z}+\frac{\partial^{2}}{\partial\rho^{2}}+\frac{1}{\rho}\frac{\partial}{\partial\rho}\right)+V_{\mathrm{core}}\left(z,\rho\right) \right.
\nonumber \\  && \left.  +qE\left(t\right)\cdot z\right]\Psi
\end{eqnarray}
Our recently developed algorithm \cite{majorosi2016tdsesolve} supports the direct numerical
integration of this TDSE with Coulomb-singularities and provides fourth
order accuracy in both space and time coordinate steps. 
We show the probability density of the tunneling electron in Fig. 1.
  
\begin{figure}
\includegraphics[scale=0.2]{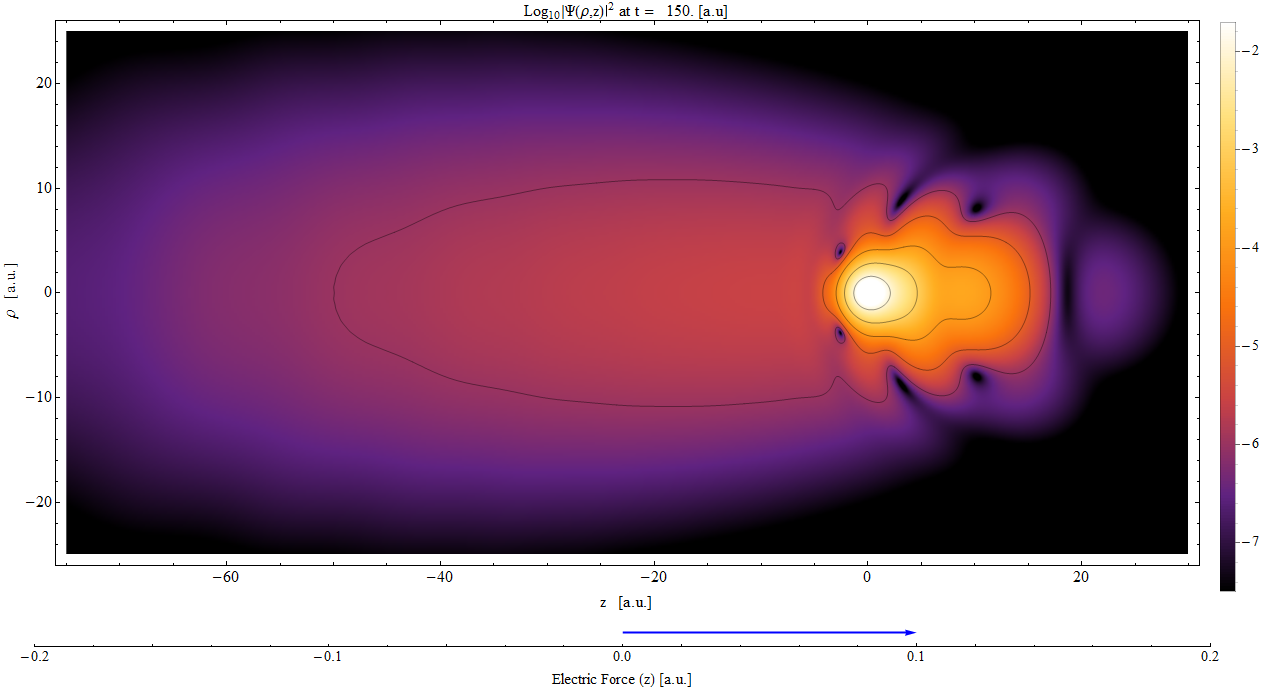}
\caption{Density plot of the base 10 logarithm of the absolute square of the wave function of a tunneling electron at the instant of the peak of the laser pulse.}
\label{fig:wavefunction}      
\end{figure}

From the numerical
solution, we first reduce the 3D wave function to the direction $z$ (by integrating over $\rho$) and then we compute the corresponding Wigner function \citep{czirjak2000ionizationwigner} as:
\begin{equation}
W(z,p_z,t)=\frac{1}{\pi}\intop_{-\infty}^{\infty}d\zeta \exp\left(2 i p_z\zeta\right)\Psi^{*}\left(z+\zeta,t\right)\Psi\left(z-\zeta,t\right).
\end{equation}
We show the Wigner function in Fig. 2 at the instant of the peak of the laser pulse.

\begin{figure}
\includegraphics[scale=0.45]{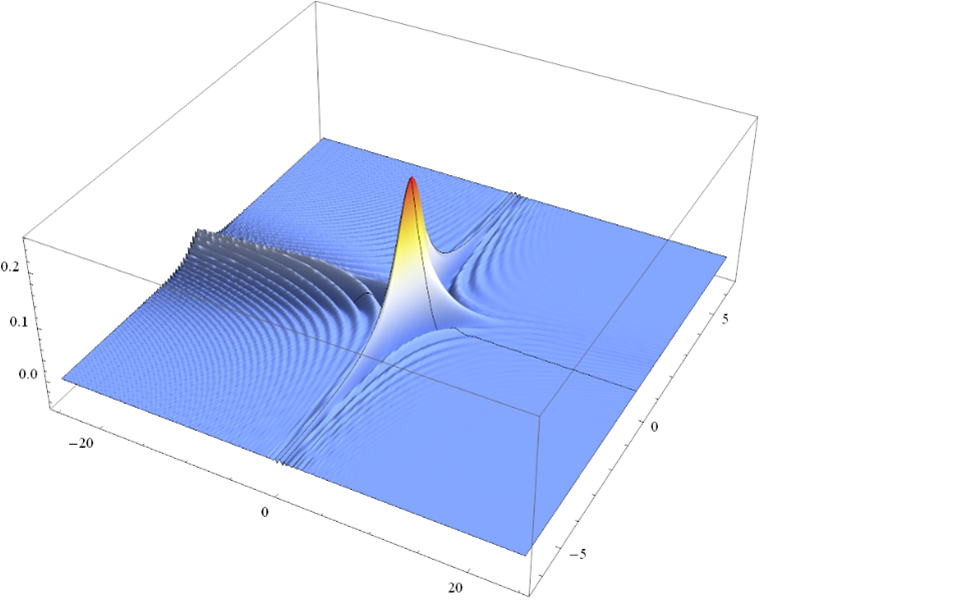}
\caption{Color coded 3D plot of the Wigner function of the tunneling electron's quantum state along the direction of the laser polarization, shown at the instant of the peak of the laser pulse.}
\label{fig:Wigner3D}      
\end{figure}

Based on the moments of the Wigner function 
\begin{equation}
P_{n}\left(z,t\right)=\int p_z^{n}W(z,p_z,t)dp_z,
\end{equation}
it is intuitive to obtain
the quantum momentum function 
\begin{equation}
\bar{p_z}\left(z,t\right)=\frac{P_{1}\left(z,t\right)}{P_{0}\left(z,t\right)}
\end{equation}
which we next compare with classical trajectories.

\section*{Results}

\begin{figure}
\includegraphics[scale=0.45]{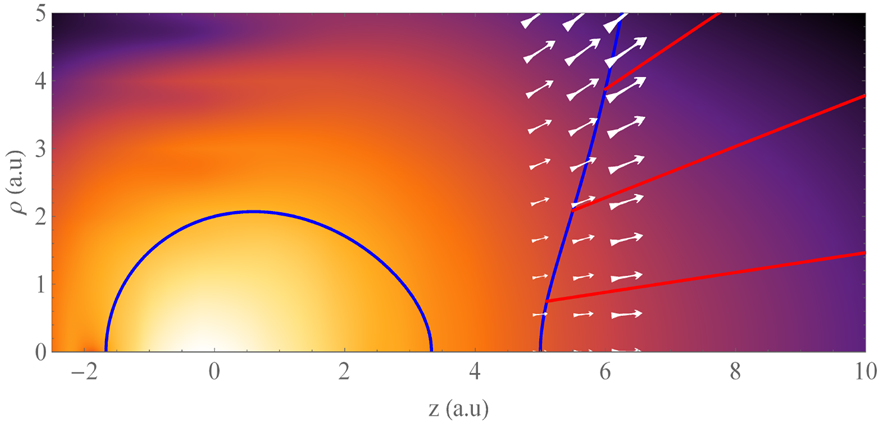}
\caption{Plot of the real space electrons trajectories (red) starting with exit momenta (white arrows) based on the 3D quantum momentum function, on top of the density plot of the quantum probability density at the laser peak power. We plot the contours of the tunnel region in blue.
}
\label{fig:init}      
\end{figure}

In order to compare the classical and quantum dynamics of the tunnel
ionization process, we compute the classical electron trajectories
starting from the tunnel exit with non-zero exit momentum, given by
the quantum momentum function, see Fig. \ref{fig:init}. 
The motion is governed by the laser
pulse according to the Newton-Lorentz equation:

\[
\mathbf{\ddot{r}}=-q(\mathbf{E}\left(t\right)+\dot{\mathbf{r}}\left(t\right)\times\mathbf{B}\left(t\right)),
\]

\begin{figure}
\includegraphics[scale=0.45]{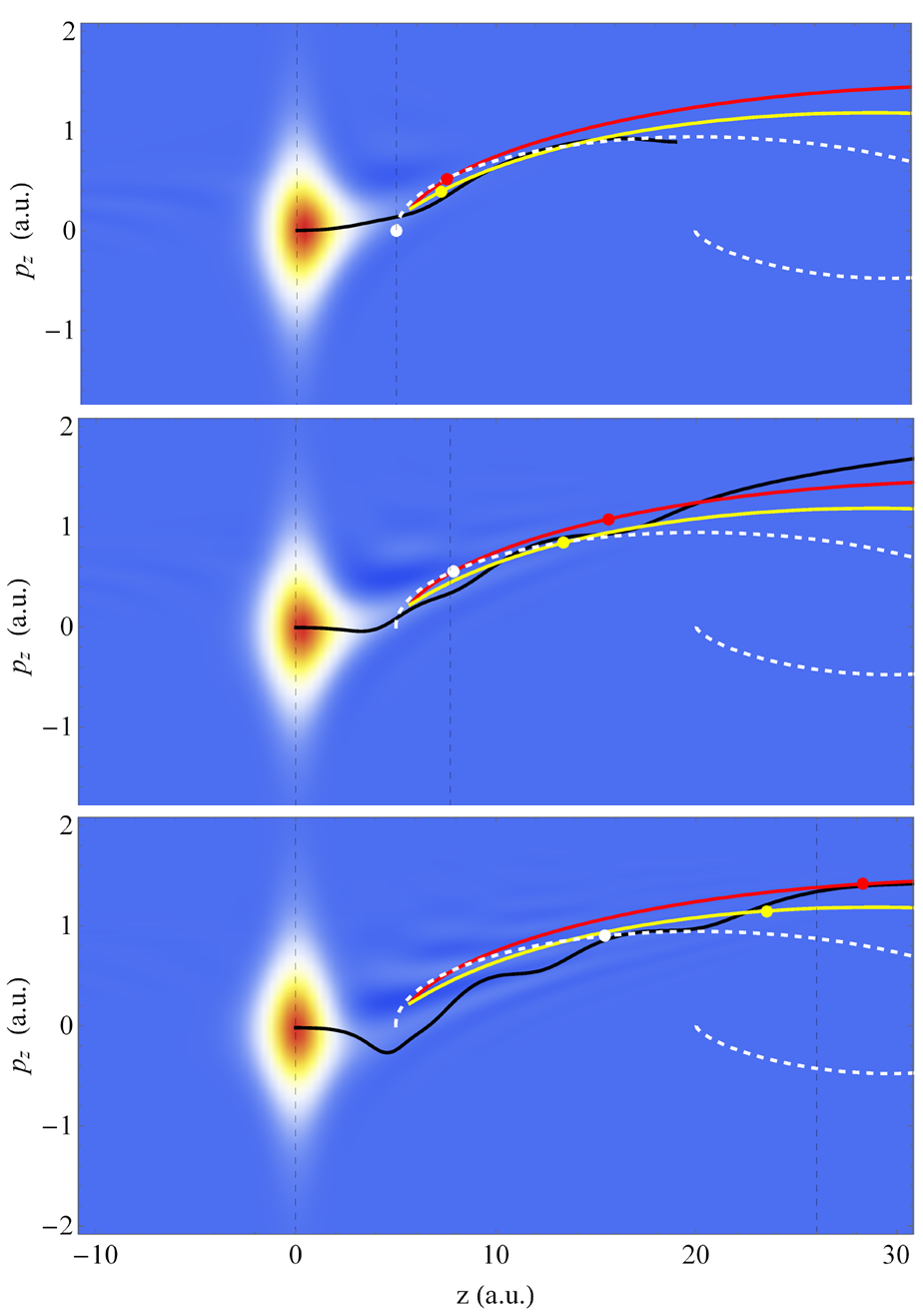}
\caption{Time evolution of the phase space dynamics of an electron with tunnel exit time of 145 a.u., shown at time instants of 150, 160 and 170 a.u. We plot the classical trajectories of the simple-man model (white dased), the improved trajectories with non-zero exit momentum with (yellow) and without (red) Coulomb-interaction, in comparison with the quantum momentum function (black) on top of the density plot of the Wigner function. Vertical dashed lines: positions of the ion core and tunnel exit.
}
\label{fig:Wigner_and_trajectories}      
\end{figure}

We compare the resulting classical motion along the direction of the laser polarization to the corresponding quantum dynamics with the help of phase space trajectories and the Wigner function in Fig. \ref{fig:Wigner_and_trajectories}.

Using the analytical solution of the Newton-Lorenz equation above,
we obtain the following equations which we use to reconstruct approximately
the tunnel exit time $\left(t_{i}\right)$ and the exit momentum components
($p_{z}^{0}$ and $p_{\rho}^{0}$), based on the momentum components
measured by a (e.g. TOF) detector ($p_{z}^{d}$ and $p_{\rho}^{d}$):
\begin{align}
p_{z}^{0} & =p_{z}^{d}\cos\left(\beta\left(t_{i}\right)\right)+\left(c-p_{\rho}^{d}\right)\sin\left(\beta\left(t_{i}\right)\right)\\
p_{\rho}^{0} & =c-\left(c-p_{\rho}^{d}\right)\cos\left(\beta\left(t_{i}\right)\right)+p_{z}^{d}\sin\left(\beta\left(t_{i}\right)\right),\\
\beta\left(t_{i}\right) & =\frac{E_{0}}{16c\omega}\left[6\sin\left(\frac{2}{3}\omega t_{i}\right)-8\sin\left(\omega t_{i}\right)+3\sin\left(\frac{4}{3}\omega t_{i}\right)\right].
\end{align}
Based on our simulations, the relative errors for the reconstructed
tunnel exit time and for the exit momentum component perpendicular
to the laser polarization ($p_{\rho}^{0}$) are below 0.1\%, and for
the exit momentum component parallel to the laser polarization ($p_{z}^{0}$)
below a few percent.

By detecting all those electrons which ultimately leave the atom,
we can reconstruct the last tunnel exit time for such electrons, which
is then also the start time for those electrons which can recombine
and thus contribute photons to the resulting isolated attosecond pulse.

\section*{Conclusions}

We present some improvements to the description of optical tunneling
and we show how to associate tunnel exit time and exit momentum to
the classical trajectories based on the quantum momentum function.
The resulting electron trajectories fit very well to the quantum description
given by the Wigner function and by the quantum momentum function.
Our results also show that the tunnel ionization starts a few atomic
time units before the main peak of the linearly polarized single-cycle
laser pulse. We also derived an approximate analytic formula to reconstruct
the exit momentum and ionization time from electron momentum data
measured with a usual time-of-flight electron detector. This also
allows to determine an improved start time for those electrons that
can recombine with the parent ion, which then may help to determine
the ''time zero'' of the attosecond
pulse.

\bibliographystyle{apsrev4-1}
\bibliography{manuscript}

\end{document}